\documentclass{llncs}
\usepackage{listings}
\usepackage{enumerate}
\usepackage{multirow}
\usepackage{url}
\usepackage{color}
\usepackage{todonotes}
\usepackage{amsmath}
\usepackage{amssymb}
\usepackage{array}

\usepackage{semantic}

\usepackage{graphicx}
\graphicspath{}
\usepackage{subfigure}

\usepackage{macroses}
\usepackage{float}

\usepackage{algorithm}
\usepackage{algorithmicx}
\usepackage{algpseudocode}

\algblockdefx[mydo]{MyDo}{EndMyDo}%
    [1]{\textbf{do} #1}%
    {\textbf{end do}}

\algblockdefx[forall]{ForAll}{EndForAll}%
    [1]{\textbf{for all} #1 \textbf{do}}%
    {\textbf{end forall}}

\floatstyle{ruled}
\newfloat{protocol}{thp}{lop}
\floatname{protocol}{Protocol}

\graphicspath{{figures/}}

\begin{document}
\title{How to Generate Security Cameras: Towards Defence Generation for Socio-Technical Systems}

\author{Olga Gadyatskaya}

%\inst{1}

\institute{SnT, University of Luxembourg\\
olga.gadyatskaya@uni.lu}

\maketitle              % typeset the title of the contribution

\begin{abstract}
Recently security researchers have started to look into automated generation of attack trees from socio-technical system models. The obvious next step in this trend of automated risk analysis is automating the selection of security controls to treat the detected threats. However, the existing socio-technical models are too abstract to represent all security controls recommended by practitioners and standards. In this paper we propose an attack-defence model, consisting of a set of attack-defence bundles, to be generated and maintained with the socio-technical model. The attack-defence bundles can be used to synthesise attack-defence trees directly from the model to offer basic attack-defence analysis, but also they can be used to select and maintain the security controls that cannot be handled by the model itself.

Full version of this paper has appeared in GraMSec 2015, to be published by Springer.

\end{abstract}

\section{Introduction}\label{sec:introduction}

Models are used in all stages of the security process: from security requirements elicitation and organisational risk assessment to run-time verification and business process compliance audit. Often these models are inter-connected. For example, if a security requirements model for a software system was elicited, on the later stage it may be re-used to design the security testing process for this system. At the same time, as manual production of security models is very tedious and error-prone, many researchers and practitioners look into automating the model creation and transformation processes.

Recently security researchers have looked at systematic design \cite{Paul-GraMSec-2014} and automated generation of attack models \cite{Ivanova-Generation-2014}, \cite{D3.4.1-2014}, \cite{Ou-CCS-2006}, \cite{Vigo-CSF-2014}, such as attack graphs and attack trees, from system models. This model transformation allows to switch the view from the system description perspective to a compact representation of possible attacker actions. At the same time, given the generated attack model, the system defender is interested to find the weakest links: the spots in the model where additional security controls can be introduced to improve protection and eliminate potential attacks. Therefore, automated generation of defences is an obvious next step in the process. 

In this paper we look at socio-technical models as succinct abstractions of large organisations. Such models capture simultaneously locations, actors and objects in the system. They often take into account both physical and digital domains and offer to a human analyst the means to represent ``the world as it is''. That means that the designer of socio-technical systems does not need to be a security or risk analysis expert. She only needs to know the intricacies of her own company (department) to be able to model it. With the system model at hand, at the next step the attack generation tools aim at automatic creation of attack scenarios that can be further discussed by security professionals. The overall idea of this process is to automate threat scenarios identification (an important aspect of risk analysis) as much as possible. 

In this paper we would like to push the envelope even further. Our main question is: \emph{given a socio-technical system model, how to find and capture, possibly automatically, the security controls that will counteract the discovered threats?} Indeed, the main goal of risk analysis is to improve the existing system by introducing new security controls, so that the most dangerous or easily executed attacks are thwarted. Therefore, automated creation of attack scenarios only is not yet a full solution. 

We want to look at perspectives and limitations of automated defence generation from socio-technical models. It seems that the main obstacle to rich defensive strategies generation directly from the model is the fact that socio-technical models do not capture many security controls.  

To find an answer to the main question, we start from investigating the security controls (defences) already present in an advanced socio-technical model and propose a scheme to extract these controls, together with the attack steps, in the compact format of attack-defence bundles. We then evaluate the limitations of the extracted defences inherent from the socio-technical model and discuss how to overcome these limitations. We argue that an attack-defence model needs to be maintained (in parallel with the socio-technical model) that can capture not only the attacker's view but also the defender's view of the system. In this paper we have chosen attack-defence trees \cite{Kordy-JLC-2014} as the basis for the attack-defence model. As an alternative to this model, one can choose, for example, attack-countermeasure trees \cite{Roy-SCN-2011}.

The goal of this paper is to propose an attack-defence view for socio-technical models that can capture simultaneously attacker's options and available/proposed countermeasures in the system. The main idea is that given that view it can be easily synchronised with the model (but it contains richer defence information than the model), and it can be used to synthesise attack-defence trees and evaluate different interesting attributes. 

%%%%%%%%%%%%%%%%%%%%%%%%%%%%%%%%%%
\section{Socio-Technical Models versus Attack-Defence Models}\label{sec:motivation}
%%%%%%%%%%%%%%%%%%%%%%%%%%%%%%%%%%

As socio-technical models are abstractions, they do not capture all defensive mechanisms that can be available in an organisation, but only a subset of them. Indeed, it is impossible to model all security-relevant devices, protocols and behaviours in a single model. Typically, socio-technical models look at capturing organisational infrastructure (e.g., \cite{D1.3.1}, \cite{ANKH}, \cite{Exasym-2008}, \cite{Lenzini-2015}), but sometimes they can focus only on some aspects of human-computer interactions (e.g., \cite{Radomirovic-2015}, \cite{STEAL-2014}).

Since all security aspects cannot be captured by a socio-technical model without overcomplicating it, we argue that there is a need to maintain a separate view of attack and defence capabilities of the system together with the socio-technical model. Preferably, we should be able to trace the objects in the socio-technical model into the attack-defence model and back. 

\textbf{Requirements for the attack-defence model}

The first requirement for the chosen attack-defence model is that \emph{the defences that are already captured by the model need to be represented explicitly in the attack-defence model}. Indeed, we would like to faithfully represent the system security state. So, if some security control is captured by the system model, it should be translated into the generated attack-defence model. 

Secondly, we want to propose a way to \emph{update the generated defender's view (the security controls obtained directly from the system model) with more security controls and countermeasures} of the organisation. This \emph{update needs to be consistent}: once a security control is captured in the attack-defence model, it should be traced to an object in the system model. For example, if our approach identifies that a security camera is to be placed in a certain location in the system, all attack scenarios that involve that location should be updated to take the camera into account. In this way later on one can investigate automated defence generation process that will maintain consistency of the socio-technical system. 

%%%%%%%%%%%%%%%%%%%%%%%

\textbf{Background}

In this paper we use the TREsPASS socio-technical model \cite{D1.3.1} that is graph-based. We can briefly summarise this model as follows. Locations in the system represent physical and network locations; actors model humans and processes; and items can be physical or digital objects. Edges among locations represent connectedness (e.g., adjacent rooms), and all actors and items are located somewhere in the system. Actors can possess items, and items can be embedded into other items. Some locations have access control policies attached to them. These policies specify a set of credentials (items in the system) an actor needs to possess to enter the location or access the object. These policies can also be formalized by more complex predicates capturing, e.g., role-based access control or trust relationships among actors. 

As the starting point for the attack-defence model, we consider the process of attack trees generation by policy invalidation that relies on structural information about the system \cite{Ivanova-Generation-2014}, \cite{D3.4.1-2014}, \cite{Kammuller-2013}. This process was initially designed for the TREsPASS socio-technical model \cite{D1.3.1}, but it can be applied to other socio-technical models capturing systems as graphs, e.g., \cite{Exasym-2008}, \cite{ANKH}, \cite{Portunes-2010}, \cite{Lenzini-2015}, because it is reachability-based.

In short, this process is started by choosing an asset among the entities in the system. The attacker is also selected among actors in the system (the main goal of the attacker is to invalidate the security policy, e.g., confidentiality or integrity policy, associated with this asset). Then, based on the reachability reasoning, the process systematically searches for the ways for the attacker to access the asset. For example, consider the asset to be a sensitive document located in a locker in the manager's office, and the attacker to be an insider (an employee) working on the same floor. To access the document, the attacker can try to access the locker and open it (an AND-decomposition \cite{Kordy-JLC-2014}). This might require possession of the key to the locker that needs to be obtained elsewhere in the system.  Alternatively (an OR-decomposition with the previous attack), also the manager has access to the locker and the document. Thus, the attacker can get access the document by influencing the manager. This can be implemented through, e.g, social engineering (for instance, befriending the manager, or hiring an external actor to pretend to be a higher executive who needs the document), bribing, or coercing the manager.

In this small motivating example we see that two general attack strategies come into play: the attacker can actively pursue moving across the system and collecting items that will open him the way to the desired asset, or the attacker can attempt to orchestrate actions of other actors in the model so that they will do the necessary actions for him. Irrespectively of the chosen strategy, the process of attack trees generation by policy invalidation will systematically identify available (reachable) steps, add them to the tree, and refine those steps further, producing a complete attack tree in the end \cite{Ivanova-Generation-2014}. Notice, that this summary is a simplification of the overall process, and we encourage the reader to refer to the original articles about the approach for more details \cite{Ivanova-Generation-2014}, \cite{D3.4.1-2014}, \cite{Kammuller-2013}.

\section{Attack-Defence Model}\label{sec:extraction}

\textbf{Extraction of defences from the model}\\
The only security controls the TREsPASS socio-technical model captures are access control policies that restrict access to certain locations. These policies can correspond to physical (locks) or digital (password check) means (policy enforcement mechanisms) implemented in the system to restrict access to assets. Therefore, we propose to make explicit in the attack generation process the fact that the attacker needs to overcome the restrictions imposed by security policies. To achieve that we will use attack-defence bundles that are based on the attack-defence tree formalism \cite{Kordy-JLC-2014}. 

Intuitively, the attacker can chose from two approaches to deal with security policies in the system. He can attempt to satisfy the access control policy (for example, by collecting the necessary credentials or coercing someone with the right credentials) or he can try to circumvent the policy (e.g., by forcing the lock). The first approach is in line with the attack tree generation by policy invalidation process, because it can be automatically designed based on reachability. If we want to refine the second approach, we need to understand how exactly different policies (more precisely -- enforcement mechanisms for these policies) can be circumvented. There is a need to represent the human expert knowledge in circumventing different security mechanism in such a way that it is useful for automated generation process. To achieve that, one can use, for example, the hierarchical approach to attack representation suggested in \cite{Pinchinat-WFMDS-2014}.

Indeed, the enforcement mechanisms for access control policies defined in the socio-technical model can be automatically introduced into attack-defence trees. If the knowledge about breaking certain kinds of enforcement mechanisms is available in a suitable format (e.g., the hierarchical representation), then the attack-defence trees can be further refined based on that information. Further analysis based on the attack-defence trees produced at this stage (e.g., computation of the most probable or the most cheap attack for the attacker \cite{Bagnato-IJSSE-2012}) can identify the missing enforcement mechanisms. For example, if in the sensitive document scenario the attacker can directly access the document because the locker does not require any key (no access control is enforced for the document), it might be the first recommendation for improving security of the organisation: to introduce some appropriate access control mechanism (e.g., an actual lock with the key) to protect access to the document.

\subsection{Simplified Socio-Technical Model}
We introduce a simplified TREsPASS socio-technical model to exemplify the attack-defence model creation. The simplified model allows to reason only about potential reachability. However, this is already very useful for risk analysis, as quantitative evaluation of the possibility that an attacker accesses some system elements can simplify risk analysis for human analysts \cite{Othmane-CS-2015}.

The simplified model captures simultaneously organisation's infrastructure topology for both physical and digital locations, as well as actors moving around this infrastructure (these can be persons or processes).  In the model these entities are represented as a set of model elements $N$ that is a union of a set of infrastructure locations $N_i$, actors $N_a$, and objects $N_o$. We consider two domains: $Ph$ is the physical space (model elements in this domain are physical entities, including, e.g.\ rooms, persons, and items), while $Dg$ is the digital space (network locations and processes are in this domain), such that $N = Ph \cup Dg$, and $Ph \cap Dg = \emptyset$.

Some model elements are connected. We denote as $E \subseteq N \times N$ the set of directed connections. All edges $e$ in $E$ are of the following types:
\begin{itemize}
\item $e \in E_{ii} \subseteq N_i \times N_i$: connections between infrastructure locations (rooms, corridors, etc.). These connections are assumed bi-directional. More precisely, if $(i_1,i_2) \in E_{ii}$ then $(i_2,i_1) \in E_{ii}$.
\item $e \in E_{ai} \subseteq N_a \times N_i$: placement of actors in the infrastructure; 	
\item $e \in E_{oi} \subseteq N_o \times N_i$: placement of objects in the infrastructure;
\item $e \in E_{oa} \subseteq N_o \times N_a$: placement of objects that are carried around by actors; 
\item $e \in E_{oo} \subseteq N_o \times N_o$: placement of objects that are inside other objects; here $e = (o_1, o_2)$ denotes an object $o_1$ located within an object $o_2$.
\end{itemize}
Mutual intersections of $E_{ii}, E_{ai}, E_{oi}, E_{oa}, E_{oo}$ are empty sets. Elements of the same domain can be connected liberally. However, some self-evident restrictions apply when connections between elements of the physical and digital domains are considered. For example, a data file cannot be located in an office or inside a cupboard. We allow multiple locations for the same actor and object. This corresponds to the possibility of actors to move in the model, and represents that some items can appear in several locations. 

We define a location function \loc{}: $N$ $\times$ $N$ as follows: 
$\forall n \in N $ \loc{n}: = $\{ l \in N | (n, l) \in E \}$.

Notice that for infrastructure locations or actors the function \loc{} returns infrastructure locations where these model items are accessible from. However, as objects can be accessible from actors or other objects, \loc{} may return any type of items in the model. 

\textbf{Policies}

Let $P$ be a set of policies defined in the model. We consider access control policies represented as tuples restricting access to element $n$. The \emph{local policy} $\delta_n$ is a set of individual access control configurations. Each access control configuration $p \in \delta_n$ is a tuple $\langle Cred, atLocation, EM\rangle $, where $Cred \subseteq N_o$ is a set of credentials required to get access, $atLocation \in N$ s.t. $(n, atLocation) \in E$ is a model element from which access to $n$ is granted (e.g. access from the office to the locker is granted with the key in the example in Sec. \ref{sec:motivation}), and $EM \in N$ is a reference to the mechanism enabled in the model to enforce the policy. $EM$ can be the same as $atLocation$, meaning that the enforcement mechanism is implemented right at the spot (e.g., a lock), it can be an actor (e.g., a security guard checking identity documents or a process implementing access control), or an object. Notice that we assume that $c \in Cred \subseteq N_o$ is an asset present in the model, which can be either an item or data. 

In theory, different access control configurations of the same local policy $\delta_n$ can be enforced by different enforcement mechanisms. For example, to access a building employees might use a badge applying it to an RFID-reader, or they might show their IDs to a security guard.

\subsection{AD-Bundles Generation}\label{subsec:generation}

%%%%%%%%%%%%%%%%%%%%%%%%%%%%%%%%%%%%%%%%%%%%%%%%%%%%%%%%%%%%%%%%%%%%%%%%%%%%%%%%%%%%%%

We will now show how to generate attack-defence bundles (AD-bundles) that can be used to capture the attack-defence state of the system.
AD-bundles are generated for individual assets. They consist of attack nodes that correspond to gaining access to items in the model and attacking these items, and defence nodes that represent protections offered by the local policies in place. Notice that the bundles are attacker-agnostic, and they refer only to the system configuration regarding some particular item. Our notation abuses the standard notation for attack-defence trees, as we use AD-terms to represent both the tree structure and to refer to concrete attacker goals. We also define different types of AD-terms. This is syntactic sugar to ease the type representation, as types are used to put bundles together and synthesise AD-trees.

\paragraph{Attack node types.}
We consider attack nodes can be of the following types. 
\begin{itemize}
\item \taccess$_n$ is an attack node that represents that the attacker gains access to item $n$. 
\item \taccfrom$_{n,l}$ represents the goal of the attacker to access item $n$ from specific model element $l$. This node type explicitly states the way $n$ is accessed in the model, thus allowing us to understand immediately what access control policy is applicable (by looking at the $atLocation$ attribute). 
\item \tbreak$_n$ represents the goal of the attacker to somehow disable an access control mechanism implemented in $n$ (this enforcement mechanism can protect assets not located in $n$).
\item \tbreakpol$_p$ represents the goal of the attacker to overcome protection of an individual access control configuration $p$.
\item \tsatpol$_p$ represents attacker's goal to satisfy access control configuration $p$ (by collecting all necessary credentials).

\end{itemize}

\paragraph{Defence node types.}
The defence nodes can be of the following types:
\begin{itemize}
\item \tdefence$_{n,l}$ represents the defence of enforcement mechanisms enforcing policies at $l$ to control access to $n$ (notice that the enforcement mechanism itself can be located elsewhere).
\item  \tdefpolicy$_p$ represents protection offered by an individual access control configuration for some $p \in \delta_{n}$. 
\end{itemize}
Notice that term types \tbreakpol\ and \tdefpolicy\ are required to satisfy the requirement of AD trees for the unique child of the opposite type \cite{Kordy-JLC-2014}. 

\paragraph{Bundle construction.}
Let $n \in N$ be an item in the model. An AD-bundle \baccess$_n$ that characterises accessing $n$ is constructed as follows.

We start by setting the root of the bundle to \taccess$_n$, as this is the desired attacker's goal. 

Next, \taccess$_n$ is refined:\\
\taccess$_n$ := \OR$^p$ $\Big( \taccfrom_{n,l} | l \in \loc{n}  \Big)$ \ \ // \emph{$n$ can be accessed only from an adjacent element in the model. Any of these elements is suitable for the attacker}\\

If $\nexists p = \langle Cred, l, EM \rangle \in \delta_n$ then\\
 \taccfrom$_{n,l}$ := $ \taccess_{l}$ \ \ // \emph{access to $n$ from $l$ can be implemented by simply accessing $l$. No access control policy is set up to guard this connection.}\\

If $\exists p = \langle Cred, l, EM \rangle \in \delta_n$ then\\
\taccfrom$_{n,l}$ := $c^p$ $\Big( \taccess_{l}, \tdefence_{n,l} \Big)$ \ \ // \emph{access to $n$ from $l$ can be implemented by accessing $l$. However, as there is an enforcement mechanism that controls access, the defence node is also added.}\\

\tdefence$_{n,l}$ := $\AND^o \Big( \tdefpolicy_p | \forall p \in \delta_n$ s.t. $ p = \langle Cred, l, s \rangle \Big)$ \ \ // \emph{Protection of access from $l$ to $n$ is implemented via individual policy configurations.}

\tdefpolicy$_p$ := $c^p \Big( \tbreakpol_p \Big)$ \ \ // \emph{syntactic sugar to switch back to attacker's view} 

\tbreakpol$_p$ := $\OR^p \Big( \tsatpol_p, \tbreak_{s} \Big)$, where $p = \langle Cred, l, s \rangle$ \ \ // \emph{Attacker can either satisfy the individual policy configuration $p$, or he can break the enforcement mechanism $s$ that enforces this configuration $p$. }

\tsatpol$_p$ := $\AND^p \Big( \taccess_{cred} | \forall cred \in Cred \Big)$, where $p = \langle Cred, l, s \rangle$ \ \ // \emph{To satisfy the configuration the attacker needs to access all credentials in the set $Cred$ identified in this configuration}.

\begin{figure}[htb]
\centering
\includegraphics[width=0.85\columnwidth]{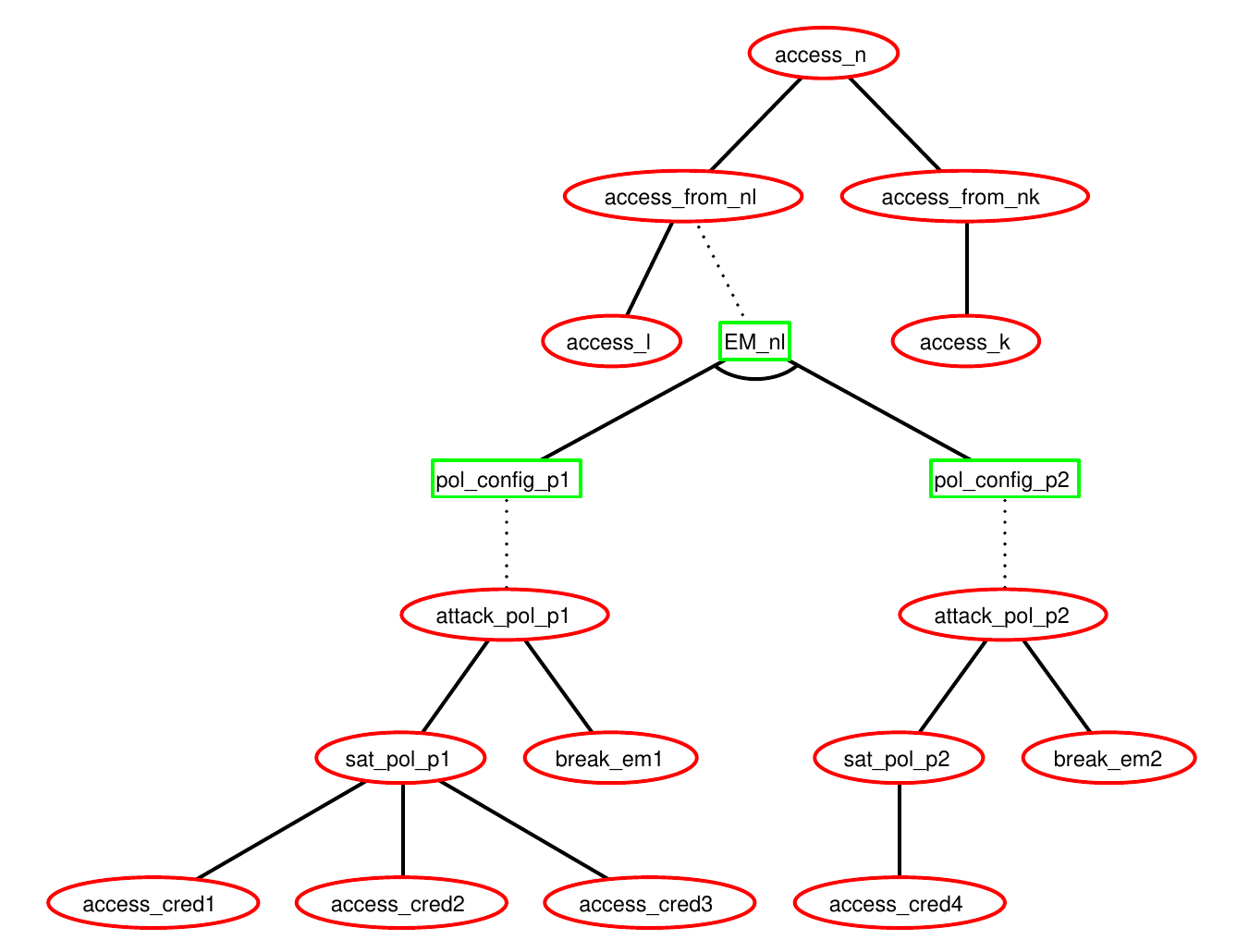}
\caption{An AD bundle.}
\label{fig:bundle}
\end{figure}

We provide an example of an AD bundle in Fig. \ref{fig:bundle}.
By construction, for each bundle $\baccess_n$ its leaf nodes are either terms of the same type (\taccess$_l$ for some $l$), or terms $\tbreak_s$. We do not refine terms of the type \tbreak$_{s}$ because the model itself lacks the knowledge how enforcement mechanisms can be broken. If an additional knowledge on breaking enforcement mechanisms will be available (e.g., as a hierarchy of attacks \cite{Pinchinat-WFMDS-2014}), this term can further expanded. 

\subsection{Approach to Synthesise AD-Trees}

AD-bundles represent attacks on individual assets in the model. They can be ``glued'' together to form AD-trees, in the spirit of attack generation by policy invalidation. In this subsection we outline an approach to synthesis of attack-defence trees.

The main requirement for AD-trees synthesis is that it should terminate. Indeed, it is easy to see that any simple loop in the infrastructure will create infinite trees if bundles are composed naively. Moreover, some bundles may appear more than once in the generated tree, creating duplicate subtrees. To avoid this, we introduce a system state that will keep track of already achieved progress and will allow to terminate the synthesis process when the attacker has achieved the goal.

\paragraph{State.}
We define now two functions that identify the state of the system. These functions will be updated as the attack tree is generated in order to keep track with the attack development.

\iffalse{
We define a boolean function \Posesses{}{} $\subseteq$ $N_a$ $\times$ $N_o$:
\begin{itemize}
\item If $(o,a) \in E_{oa}$, and  \Posesses{a}{o} := \texttt{True}.
\item If for some $o_2 \in N_o$ $(o_1,o_2) \in E_{oo}$, and \Posesses{a}{o_2} = \texttt{True}, then \Posesses{a}{o_1} := \texttt{True}.
\item Else \Posesses{a}{n} = \texttt{False}.	
\end{itemize}
Intuitively, the function \Posesses{}{} refers to objects that an actor has.
}\fi

\begin{definition}[\Reachable{}{}]	
Let \model\ = $(N,E)$ be a model. We define a boolean function \Reachable{}{} $\subseteq$ $N_a$ $\times$ $N$:
\begin{itemize}
\item If $(a,n) \in E$, \Reachable{a}{n} := \texttt{True}.
\item If for some $l \in N_i$ $(a,l) \in E_{ai}$ and $(o,l) \in E_{oi}$, then \Reachable{a}{o} := \texttt{True}. 
\item If for some $l \in N_i$ $(a,l) \in E_{ai}$ and $(a_1,l) \in E_{ai}$, then \Reachable{a}{a_1} := \texttt{True} and \Reachable{a_1}{a} := \texttt{True}. 

\item Else \Reachable{a}{n} := \texttt{False}.	
\end{itemize}
\end{definition}

This function initially captures for a given actor all items immediately reachable in the model. These items can be objects or actors located in the same location as the actor. Let \Reach{a} := \{$\forall n \in N$ s.t. \Reachable{a}{n} = \texttt{True}\}.

\begin{definition}[\Granted{}{}]
We define a boolean function \Granted{}{} $\subseteq$ $N_a$ $\times$ $N$:
\begin{itemize}
\item If for an item $n$ $\delta_n = \emptyset $ then \Granted{a}{n} := \texttt{True}.
\item If for an item $n$ there is a tuple $p = \langle Cred, atLocation\rangle$ $\in \delta_n =$ s.t. $Cred \subseteq \Reach{a} \cap N_o$ then \Granted{a}{n} := \texttt{True}.
\item Else \Granted{a}{n} := \texttt{False}.
\end{itemize}
\end{definition}

Intuitively, this function refers to some policy configuration that grants access to $n$. If \Granted{a}{n} = \texttt{True}, then there is a way for this actor to satisfy the access control policy for $n$ (possibly under condition that he arrives at the right location). 

Let us define a model state.
\begin{definition}[State]
A generated state for a model \model\ is a tuple $\langle \Reachable{}{}, \Granted{}{} \rangle$.	 
\end{definition}

\begin{definition}[\Accessible{}{}]
We define a boolean function \Accessible{}{} $\subseteq$ $N_a$ $\times$ $N$:
\begin{itemize}
\item \Accessible{a}{n} := \Reachable{a}{n} \AND\ \Granted{a}{n}
\end{itemize}

\end{definition}

\paragraph{Bootstrapping}
Given a model \model= $\langle N,E \rangle$ produced by a modeller, the functions \Reachable{}{}, \Granted{}{} and \Accessible{}{} are initially computed from \model. First we compute a transitive closure of reachable locations:
\begin{itemize}
\item \Reachable{a}{n} 	:= \Reachable{a}{n} \OR\ ($\exists l \in N:$ \Accessible{a}{l} \AND ($(l,n) \in E$ \OR $(n,l) \in E$) )
\end{itemize}
Notice that here we do not re-compute the function \Granted{}{}, and thus, eventually, the reachable objects set for each actor will increase only with locations that are not guarded by access control policy. Once \Reachable{}{} is recomputed, it can be used to quickly evaluate whether an actor can reach certain locations in the original model (where may he end up).

\subsubsection{Synthesis of AD-trees from Bundles}

We now discuss composition of generated attack-defence trees. An attack-defence tree $\tree(\eta,\alpha)$ is synthesised for a chosen attacker $\eta \in N_a$ and a target asset $\alpha \in N_o$. The root node is the bundle \taccess$_\alpha$. For each leaf node of the type \taccess$_b$ we can compute its value by referring to the corresponding AD bundle $\baccess_b$. 

\paragraph{Bundle Value.}
In the simplest case we use propositional semantics for evaluating AD-bundles and, eventually, AD-trees \cite{Kordy-JLC-2014}. For leaf nodes of the type \taccess$_n$, \taccess$_n$ $\equiv$ \Accessible{\eta}{n}. For leaf nodes of the type \tbreak$_s$, \tbreak$_s$ $\equiv$ \texttt{False} in the current synthesis approach. Thus, given a bundle for asset $n$, we can evaluate its value based on the values of the leaf nodes available. By updating the model state as attack progresses (more items become reachable to the attacker) we can eventually evaluate the target bundle, once all its descendants become evaluated. As state changes monotonically, the process will eventually terminate.

%%%%%%%%%%%%%%%%%%%%%%

\section{Introducing New Defences}\label{sec:2ndstep}
The enforcement mechanisms for access control policies are not the only type of security controls that organisations use. Moreover, access control is not the only remedy that can be advised to improve security. Indeed, the existing risk analysis standards and security catalogues that guide practitioners in risk analysis identify many types of security controls and countermeasures. Many of those (for example, security cameras) cannot be captured by socio-technical models directly, because it will introduce unnecessary complications to the model. Some countermeasures can be introduced as properties of system elements (e.g., after a security training the employees might become less susceptible to social-engineering), but not as independent elements of the system. 

We want to be able to update the attack-defence model of our system, captured by the suite of attack-defence bundles, after the first stage of automated generation. At this second stage we would like to obtain more complete attack-defence bundles with new defence nodes added that can capture additional security countermeasures (either existing in the model already or newly proposed once). We have two main questions associated with the newly introduced defences: \emph{how to generate/propose new defences} and \emph{where to place them in the attack-defence model} to keep the consistency across many attack scenarios and system updates. We start by addressing the second question first.

\textbf{Where to put new countermeasures}\\
Given an AD bundle representing the goal of an attacker to access asset $n$, two types of attack nodes are the candidates to be protected from by some countermeasures: the root node \taccess$_n$ and its children \taccfrom$_{n,l}$. Indeed, for the connectors to other bundles (the leaf nodes \taccess$_b$) it will make sense to introduce defences at the corresponding bundle to ensure the consistency requirement. For the nodes \tsatpol$_p$, the attacker's goal is to satisfy the policy by finding the right credentials. It is not obvious what can be done as a protective measure besides protecting the credentials themselves. As for the nodes representing circumventing the enforcement mechanism, \tbreak$_{s}$, we do not have enough details for the moment how the attacker is going to break it. If this node is to be refined using some attack pattern library, it is better to create a separate AD bundle for treating the scenarios and assign defences there.

Now we have candidate attack nodes to be assigned countermeasures. To select the countermeasures that could be assigned, we first review existing types of security controls. It is well-established in the security industry to classify controls as preventive, detective and corrective \cite{NIST-800-30}:
\begin{itemize}
\item \emph{Preventive}	controls focus on preventing security incidents from occurring. 
\item \emph{Detective} controls focus on detecting occurrences of security incidents.
\item \emph{Corrective} controls focus on aiding the organisation to recover from a security incident. 
\end{itemize}

From the implementation perspective, it is traditional to divide controls into the following categories \cite{NIST-800-30}:
\begin{itemize}
\item \emph{Technical} controls that are implemented typically as software controls.
\item \emph{Management}, or administrative, controls that are implemented as procedures and guidelines.
\item \emph{Operational} controls that focus on ensuring security and dependability of operations. These controls include physical security controls (physical access control, fire and water damage protection, etc.) and some controls that are difficult to classify as fully technical or physical (e.g., protection of personal computers).   	
\end{itemize}

\begin{figure}[htb]
\centering
\includegraphics[width=0.6\columnwidth]{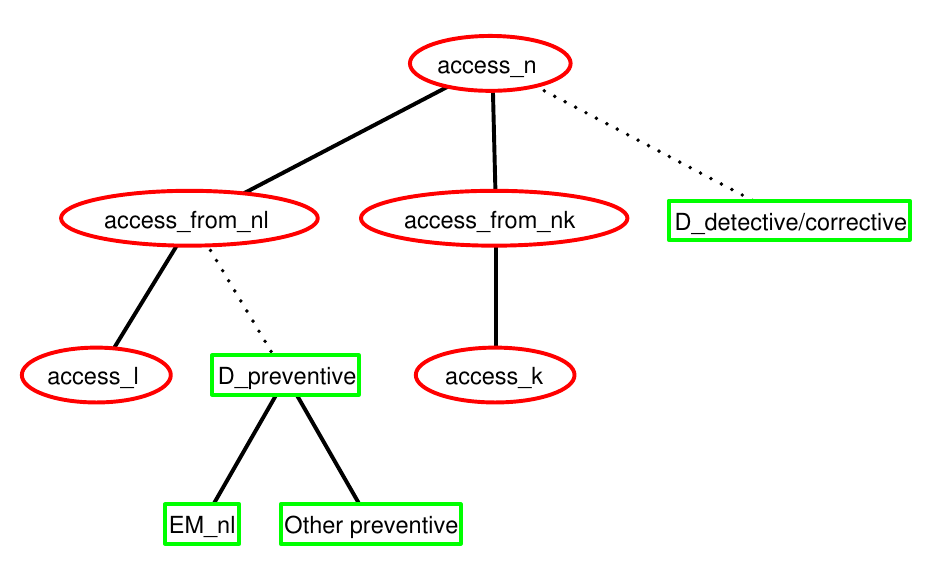}
\caption{An updated AD bundle with defence nodes in the designated positions (children of EM$_(n,l)$ not shown).}
\label{fig:updatedbundle}
\end{figure}

From this classification, we propose a way to update AD bundles with security controls in a consistent manner. The preventive controls can be added as children to the attack nodes \taccfrom$_{n,l}$, because they correspond to preventive measures for certain directed actions of the attacker. Access control policies present in the model are already embedded in the bundles at this position. To satisfy the attack-defence trees requirement of only one child of the opposite type, we will modify the bundle as in Fig.~\ref{fig:updatedbundle} (now the node \tpreventive$_{n,l}$ is a parent of the node \tdefence$_{n,l}$). 

The detective and corrective measures can be added as children to the root node \taccess$_n$ (see Fig.~\ref{fig:updatedbundle}). In this position the defence nodes are directly linked to the system object in question, be it a location, a person, or an object. The semantics of the controls placed in this position are clear: assuming the attacker has already gained access to his target, is this detectable or what can be the remedy for this? Notice that some controls in practice can be both detective and preventive (e.g., security guards). In this case, it is safe to classify them as preventive controls.

\textbf{What defences to choose}\\
The choice of security controls is a tough question in practice. Not only it requires the human analyst to know possible attacks and countermeasures, but also the analyst needs to solve a complex multi-parameter optimisation problem. Indeed, the controls addressing the same threat can have different cost, efficiency and effectiveness. They can be more or less compliant with the industry standards and best practices. They can be more or less easy to implement and easy to verify. Finally, they can be more or less desired by the organisation because of personal views of the top-management. Thus, if we just consider the baseline controls listed by NIST Special Publication 800-53 \cite{NIST-800-53}, and try to evaluate all the above-mentioned parameters in order to fully automate the defence selection (now that we know where to place them), we already will face a very complex problem. Moreover, a single mistake in evaluation of some of the values will likely make the full analysis invalid. Therefore, it is likely that human-assisted control selection cannot be fully replaced by automated defence generation, at least for some time.

Yet, we can try to facilitate the defence selection problem by further categorising the security controls based on applicability to the scenarios in question and usefulness in attribute-based computations. 

In the socio-technical model we have clear categories of objects: locations, actors and items that can belong to either physical or digital space. Thus controls can be chosen based already on simple considerations such as ``access to digital objects by processes can be protected by using technical preventive controls'', or ``access of humans to humans can be protected by administrative and physical preventive controls''. Table~\ref{tab:preventive} summarises these choices of controls. In this table, it is expected that respective controls will be introduced in the dedicated AD bundles (following the template in Fig.~\ref{fig:updatedbundle}).

\begin{table}[t!]
\centering
%\scriptsize
\caption{Controls selection based on system elements. }
\label{tab:preventive}
\scriptsize
\begin{tabular}{|c|c|c|}
\hline
\textbf{Entity}  &  \textbf{Physical space}  & \textbf{Digital space}   \\
\hline
\hline
\multicolumn{3}{|c|}{\textbf{Preventive}}\\
\hline
Location & Physical access control  & Technical access control, firewall  \\
\hline
Actor & Physical access control, Security trainings, Email filter  & Technical access control and authentication\\
\hline
Object & Physical access control  & Technical access control \\
%\cline{2-4}
\hline
\multicolumn{3}{|c|}{\textbf{Detective}}\\
\hline
Location & \multirow{3}{*}{ Security cameras, visitor logs}  & \multirow{3}{*}{System logs, IDS} \\
Actor & &\\
Object & & \\

\hline
\multicolumn{3}{|c|}{\textbf{Corrective}}\\
\hline
Location & 
Insurance, liability limitation,   & Insurance, liability limitation, \\
Actor &  business continuity plan & secure state restoring mechanisms, \\
Object &   & business continuity plan \\
\hline

\end{tabular}
\end{table}

Furthermore, following the investigation of attribute decoration on attack-defence trees by Bagnato et al. \cite{Bagnato-IJSSE-2012}, we can look at what controls contribute to computations of certain attributes. For instance, if the analyst is interested in the probability of an attack to succeed, the minimal cost of attack for an attacker, or time of executing an attack, then (under the assumption that detection cannot stop the attack) she would like to look at her preventive measures. If she is interested in the impact the attack has on her organisation (how business continuity is affected after the attack was executed), she would like to consider the preventive and corrective controls, especially the latter ones, because these ensure business continuity. Thus, if she is interested only in some attributes, the computation on AD bundles does not need to take into account all controls at once. Table~\ref{tab:attributes} summarises the control types that are the most relevant for some selected attributes. 

\begin{table}[t!]
\centering
%\scriptsize
\caption{Relevant controls for example attributes}
\label{tab:attributes}
\scriptsize
\begin{tabular}{|c|c|c|c|}
\hline
\textbf{Attribute}  &  \textbf{Preventive}  & \textbf{Detective}  & \textbf{Corrective}  \\
\hline
\hline
Risk of detection &  & \checkmark & \\
Cost of attack (for attacker) & \checkmark &  & \\
Probability of attack success & \checkmark &  & \\
Time of attack & \checkmark &  & \\
Impact of attack & \checkmark & \checkmark & \checkmark\\

%\cline{2-4}

\hline

\end{tabular}
\end{table}

Notice that the controls added at this stage will probably not follow the AD bundle notation, but will be expressed in the natural language (e.g, \emph{security training} or \emph{ID check}). This is understandable, because, as we have mentioned, models are not rich enough by nature. Yet, this is acceptable for the format, because the attack-defence model does not need to be fully formal. On the contrary, it is used to assist the human analyst to create and maintain the attack-defence view on the system. The only requirement that we have for it is the consistency, which is ensured by adding each control only to the bundle representing the attack-defence view of a particular (unique) entity in the model. Some controls can require a notion of \emph{perimeter} to be defined in the model, so that they can be uniquely assigned to the bundle corresponding to the perimeter, and not to each entity belonging to that perimeter. This is easily implementable in any socio-technical model.

\section{Related Work}\label{sec:discussion}
The question of attack trees generation from system models has been tackled in \cite{Ivanova-Generation-2014}. Similarly, \cite{Vigo-CSF-2014} and \cite{Ou-CCS-2006} worked on generating attack models from a system model. While we follow the same approach for  attacker's view, our main focus is on keeping both attacker's and defender's views consistent with the main socio-technical model.  

Attack-countermeasure trees (ACTs) is an alternative model to attack-defence trees in keeping both views simultaneously \cite{Roy-SCN-2011}. In \cite{Roy-DSN-2012} the authors have investigated optimal countermeasure selection for ACTs when a set of possible countermeasures to be implemented is already predefined. It will be interesting to investigate ACTs suitability for the attack-defence model, because they support explicit detection and mitigation countermeasure nodes (but not corrective). 

In \cite{Lenzini-2015} the authors work on directly applying model checking to a socio-technical model in order to evaluate some reachability-based security properties. 

Ferreira et al. \cite{STEAL-2014} have discussed defences suggestion in the context of the socio-technical model STEAL. They propose to apply defences at the technical and social levels of the system, what is in line with our proposal for applying security control categories in selecting defences.

\section{Conclusion and the Next Steps}

In this work we have approached the question of creating and maintaining the security controls representation in parallel to the socio-technical model. Our solution creates a set of attack-defence bundles (small attack-defence trees) that can be maintained with a socio-technical model as its separate view. The bundles are generated from the model in the beginning, but afterwards they are enriched consistently alongside the new security controls identified by a human analyst. We have also discussed how new controls can be selected based on the model entities and the attributes of interest to the analyst. This work attempts to bridge the gap between the approach of automated attack generation from system model and the manual security control selection in the traditional risk analysis. The next step is to look into the compositional attack-defence tree synthesis for more complex attribute domains.  After that, it will be possible to investigate optimal countermeasure selection, based, e.g., on the approaches suggested in \cite{Aslanyan-POST-2015} and \cite{Roy-DSN-2012}. Another further research direction is practical validation of the proposed approach on realistic case studies and evaluation of its usefulness and scalability.

\subsubsection*{Acknowledgements}
This work was partially supported by the European Commission through the FP7 project TREsPASS (grant agreement n. 318003) and by Fonds National de la Recherche Luxembourg through the ADT2P project (grant n. C13/IS/5809105).

% ---- Bibliography ----
%
\bibliographystyle{splncs03}
\bibliography{reference}

\end{document}